# Multifocal diffractive lens generating several fixed foci at different design wavelengths

LEONID L. DOSKOLOVICH,[1,2,*] EVGENI A. BEZUS,[1,2] ANDREY A. MOROZOV,[1,2] VLADIMIR OSIPOV,[3] JAMES S. WOLFFSOHN,[3] AND BORIS CHICHKOV[4]

[1]*Image Processing Systems Institute — Branch of the Federal Scientific Research Centre "Crystallography and Photonics" of Russian Academy of Sciences, 151 Molodogvardeyskaya st., Samara 443001, Russian Federation*
[2]*Samara National Research University, 34 Moskovskoe shosse, Samara, 443086, Russian Federation*
[3]*Aston University, Aston Triangle, B4 7ET, Birmingham, UK*
[4]*Laser Zentrum Hannover e.V., Hannover D-30419, Germany*
*\*leonid@smr.ru*

**Abstract:** We propose a method for designing multifocal diffractive lenses generating prescribed sets of foci with fixed positions at several different wavelengths. The method is based on minimizing the difference between the complex amplitudes of the beams generated by the lens microrelief at the design wavelengths, and the functions of complex transmission of multifocal lenses calculated for these wavelengths. As an example, a zone plate generating three fixed foci at three different wavelengths was designed, fabricated and experimentally investigated. The proof-of-concept experimental results confirm the formation of foci with fixed positions at the design wavelengths. The obtained results may find applications in the design and fabrication of novel multifocal contact and intraocular lenses with reduced chromatic effects.

## References and links

## 1. Introduction

A diffractive multifocal lens enables focusing an incident beam to several focal points along the optical axis, providing a required energy distribution between the foci [1, 2]. Such multifocal lenses are important in ophthalmic applications, optical sensors and microscopy systems [3]. In particular, contact and intraocular lenses (IOLs) that provide clear vision at near, intermediate and far distances are usually based on multifocal diffractive lenses, which create a prescribed set of foci corresponding to different vision distances [2, 4–9]. The multifocal properties are achieved either by dividing the IOL into different zones that provide refraction into different foci [4–7] or by combining a refractive lens of fixed focal length with a diffractive zone plate, which generates additional foci in several diffraction orders [2, 3, 6–10]. Until recently, the majority of the commercially available multifocal IOLs were designed to generate two foci at different distances. Trifocal lenses have now been introduced to the marketplace, which, in comparison with bifocal IOLs, provide a significant improvement in the intermediate vision [5]. Trifocal lenses contain a "segmented" diffractive relief consisting of alternating diffractive steps of different heights generating two foci in addition to the main focus. Implantation of a lens of this type was discussed in [11]. In general, the use of a segmented relief leads to the increase in the diffractive spot size.

In our previous works [9, 10], we designed trifocal lenses corresponding to the superposition of a conventional lens and a trifocal zone plate. In this case, the whole aperture of the element contributes to all the three generated foci thus providing the minimum achievable diffractive spot size.

The diffraction relief used in the existing multifocal IOLs provides the formation of additional foci (one or two) with a given position for only one given wavelength. When the wavelength is changed, the positions of the additional foci (as well as the energy distribution between the foci) are also changed. This may lead to a significant deterioration in the quality of the images formed at different wavelengths.

In the present work, we propose a method for designing 'spectral multifocal diffractive lenses', which generate prescribed sets of foci retaining their positions at several different wavelengths. As an example, we designed and fabricated a spectral trifocal zone plate having fixed optical powers at three design wavelengths. As a fabrication technique, direct laser writing in photoresist was used. Simulations and a proof-of-concept experiment confirm the feasibility of the proposed design and demonstrate the formation of three foci with fixed positions at three wavelengths.

## 2. Design method

The multifocal lens discussed here consists of two optical elements: a conventional refractive lens and a multifocal diffractive zone plate. However, these elements can be combined into a single multifocal lens [1, 10].

Let us first consider the calculation of a multifocal diffractive zone plate for a single given wavelength $\lambda$. A general method for designing multifocal zone plates was described in [1, 2]. In this case, the phase function of the multifocal zone plate takes the form

$$\varphi_{mf}(\rho;\lambda) = \Phi\left[\varphi_d(\rho;\lambda)\right] = \Phi\left[\mathrm{mod}_{2\pi}\left(\frac{\pi\rho^2}{\lambda f_d}\right)\right], \quad \rho \in [0, R], \tag{1}$$

where $R$ is the zone plate aperture radius, the function $\varphi_d(\rho;\lambda) = \mathrm{mod}_{2\pi}\left[\pi\rho^2/(\lambda f_d)\right]$ is the paraxial phase function of a lens with the focal length $f_d$. The function $\Phi[\varphi_d]$ in Eq. (1) describes a nonlinear transformation of the lens phase and provides generation of additional diffraction orders corresponding to spherical beams with foci $f_d/m$, $m = 0, \pm 1, \pm 2, \ldots$ [1, 2]. As a result, the foci of the multifocal lens consisting of a conventional refractive lens with the focus $f_0$ and a multifocal zone plate are given by

$$F_m = \frac{f_0 f_d}{f_d - m f_0}, \quad m = 0, \pm 1, \pm 2, \ldots. \tag{2}$$

The energy distribution between the foci is represented by the values $I_m = |c_m|^2$, where

$$c_m = \frac{1}{2\pi} \int_0^{2\pi} \exp\left(i\Phi[\xi] - im\xi\right) d\xi \tag{3}$$

are the Fourier coefficients of the function $\exp(i\Phi[\xi])$ $\left(\sum_m |c_m|^2 = 1\right)$ [1, 2].

For a binary-relief trifocal zone plate, the nonlinear transformation function can be defined in the following form [1, 2]:

$$\Phi[\xi] = \begin{cases} 0, & \xi \in [0, \pi), \\ \varphi, & \xi \in [\pi, 2\pi), \end{cases} \tag{4}$$

where $\varphi = 2\tan^{-1}(\pi/2)$. The value of $\varphi$ is chosen so that $|c_{-1}|^2 = |c_0|^2 = |c_1|^2 \approx 0.2884$. In this case, over 86% of the incident energy is concentrated in three beams with the foci $F_m$, $m = -1, 0, 1$. In the case of a continuous-relief trifocal zone plate, $\Phi[\xi]$ can be defined as a sine function [8, 9]:

$$\Phi[\xi] = 1.435 \cdot \sin(\xi). \tag{5}$$

The multiplicative constant in Eq. (5) provides a uniform energy distribution between the three foci: $|c_{-1}|^2 = |c_0|^2 = |c_1|^2 = 0.3$. Hence, if the function $\Phi[\xi]$ is defined by Eq. (5), 90% of the energy of the incident beam will be concentrated in the three desired beams with the foci $F_m$, $m = -1, 0, 1$. Note that the choice of the nonlinear transformation function $\Phi[\xi]$ can provide any required energy distribution between a preset number of foci [1, 2].

Diffractive phase elements (DPEs) are generally designed for a single given wavelength. At the same time, DPEs generating different or identical patterns for different wavelengths (usually, for two or three) are also widely known [12–19]. The design of such DPEs is in most cases based on sophisticated iterative optimization procedures.

To the best of our knowledge, the first diffractive optical elements proposed for working with several wavelengths are the so-called color separation gratings (CSGs) [12, 16]. These CSGs enable to separate three wavelengths $\lambda_{-1}$, $\lambda_0$, and $\lambda_{+1}$ related by the expression

$$\lambda_{\pm 1} = \lambda_0 \frac{N}{N \pm 1}, \qquad (6)$$

where $N$ is an integer, between the $-1^{st}$, $0^{th}$, and $+1^{st}$ diffraction orders: the incident radiation with the wavelength $\lambda_0$ is directed to the $0^{th}$ diffraction order, while the radiation with the wavelengths $\lambda_{\pm 1}$ is directed to the $\pm 1^{st}$ orders, respectively. In order to explain the method for designing the spectral multifocal zone plate intended for generation of several foci at several different wavelengths, let us first consider the CSG working principle. For the specified wavelengths $\lambda_0, \lambda_{\pm 1}$ related by Eq. (6), the CSG has $N$ equal-width steps per period, and their heights can be calculated analytically as [12]

$$h_j = \frac{\lambda_0}{n-1} j, \; j = \overline{0, N-1}, \qquad (7)$$

where $n$ is the grating refractive index at the wavelength $\lambda_0$. Disregarding the material dispersion, the relation between the grating microrelief height and the phase delay $\varphi$ generated at the wavelength $\lambda$ will be defined as follows [12, 16]:

$$\varphi(h_j; \lambda) = \frac{2\pi}{\lambda}(n-1)h_j, \; j = \overline{0, N-1}. \qquad (8)$$

According to Eq. (8), the grating of Eq. (7) generates the following phase delays for the incident plane waves with the wavelengths $\lambda_0, \lambda_{\pm 1}$:

$$\varphi(h_j; \lambda_0) = 2\pi j, \; \varphi(h_j; \lambda_{\pm 1}) = 2\pi j\left(1 \pm \frac{1}{N}\right), \; j = \overline{0, N-1}. \qquad (9)$$

Taking into account the phase $2\pi$-periodicity, the phase delays of Eq. (9) may be rewritten as

$$\varphi(h_j; \lambda_0) = 0, \; \varphi(h_j; \lambda_{\pm 1}) = \pm \frac{2\pi j}{N}, \; j = \overline{0, N-1}. \qquad (10)$$

According to Eq. (10), the phase delay for the wavelength $\lambda_0$ is equal to zero, which means that this wavelength is directed to the $0^{th}$ diffraction order. For the wavelengths $\lambda_{\pm 1}$, the phase delays in Eq. (10) correspond to the linear phase functions quantized into $N$ levels. These phase functions coincide with the phase delays introduced by quantized diffractive prisms designed for each of these wavelengths and providing the incident beam deflection by the angles corresponding to the $-1$-st (for the wavelength $\lambda_{-1}$) or $+1$-st (for the wavelength $\lambda_{+1}$) diffraction order of the grating. Thus, the presented analysis confirms that the grating of Eq. (7) separates the three wavelengths related by Eq. (6) between the $-1^{st}$, $0^{th}$, and $+1^{st}$ diffraction orders.

It is important to note that the CSG of Eq. (7) minimizes the function

$$F(h_1, ..., h_N) = \sum_{l=-1}^{1} \sum_{j=0}^{N-1} \left| \exp\left[i\varphi(h_j; \lambda_l)\right] - \exp\left(il\frac{2\pi}{N}j\right) \right|^2, \qquad (11)$$

where $\varphi(h_j; \lambda_l)$, $l = 0, \pm 1$ are the phase delays of Eq. (10) introduced by the grating steps of Eq. (7). The function in Eq. (11) describes the difference between the complex amplitudes of the beams generated by the grating at the wavelengths $\lambda_0, \lambda_{\pm 1}$ and the complex transmission functions providing the deflection of the beams to the orders $0$ and $\pm 1$. The established relation between the function (11) and the CSG of Eq. (7) suggests that a function similar to Eq. (11) can be used as a merit function for the design of diffractive phase elements generating required patterns for several different wavelengths. In a general case, this merit function will represent the difference between the complex amplitudes of the beams generated by the element at the design wavelengths and the complex transmission functions providing the formation of the required patterns for these wavelengths.

Following this approach, let us consider the calculation of the profile of a spectral multifocal zone plate (SMZP), which generates several prescribed foci at several arbitrarily chosen wavelengths $\lambda_l$, $l = 1,...,L$ [not necessarily related by any analytical formulas similar to Eq. (6)]. In this case, the SMZP profile is found by minimizing the difference between the complex amplitudes of the beams generated at the design wavelengths $\lambda_l$, and the complex transmission functions of the multifocal lenses $P_{mf}(\rho;\lambda_l) = \exp\left[i\varphi_{mf}(\rho;\lambda_l)\right]$ calculated for these wavelengths. Assume that the radial profile of the SMZP has $N$ steps with the same width $\Delta = R/N$ and the heights $h_j$, $j = \overline{0, N-1}$. Let us denote by $P_{smf}(h_j;\lambda_l) = \exp\{i(2\pi/\lambda_l)[n(\lambda_l)-1]h_j\}$ the complex transmission function of the SMZP for the design wavelength $\lambda_l$ at the point $\rho_j = (j+0.5)\Delta$, where $h_j$ is the microrelief height at $\rho_j$, and $n(\lambda_l)$ is the refractive index of the element. Then the $h_j$ values can be found by minimizing the following merit function:

$$F(h_0,...,h_{N-1}) = \sum_{l=1}^{L} w_l \sum_{j=0}^{N-1} \left| P_{smf}(h_j;\lambda_l) - P_{mf}(\rho_j;\lambda_l) \right|^2 \to \min, \quad (12)$$

where $w_l$ are the weight coefficients $\left(w_l > 0, \sum_{l=1}^{L} w_l = 1\right)$ describing the required energy distribution between the foci. According to Eq. (12), the $h_j$ values can be found independently from the following condition:

$$F_j(h_j) = \sum_{l=1}^{L} w_l \left| P_{smf}(h_j;\lambda_l) - P_{mf}(\rho_j;\lambda_l) \right|^2 \to \min. \quad (13)$$

In the SMZP calculation, intrinsic technological limitations on the microrelief height $h_{max}$ and the number of relief levels $M$ have to be taken into account. Let us assume that the heights $h_j$ can take one of the following $M$ values: $h_j \in \{0, h_{max} \cdot (1/M),..., h_{max}[1-1/M]\}$. In this case, the $h_j$ values minimizing the function of Eq. (13) can be found by brute-force search:

$$h_j = h_{max} \frac{m_j}{M}, \quad m_j = \arg\min_{m \in \{0,...,M-1\}} \left[ \sum_{l=1}^{L} w_l \left| P_{smf}\left(h_{max}\frac{m}{M};\lambda_l\right) - P_{mf}(\rho_j;\lambda_l) \right|^2 \right]. \quad (14)$$

Thus, the calculation of the SMZP is performed using Eq. (14). In order to obtain a spectral multifocal zone plate with three fixed foci $F_m$, $m = -1, 0, 1$ at the wavelengths $\lambda_l, l = 1,...,L$, complex transmission functions of the trifocal zone plates have to be used as the functions $P_{mf}(\rho;\lambda_l)$ in Eq. (14). The phase functions $\varphi_{mf}(\rho;\lambda_l)$ can be calculated from Eqs. (1) and (4) or Eqs. (1) and (5) with the same $f_d$ value.

## 3. Design examples

### 3.1. Trifocal zone plate for the three wavelengths 450, 540 and 580 nm

In order to assess the efficiency of the proposed approach, we designed a spectral trifocal zone plate defined by Eqs. (1), (5), and (14) for the following parameters: $f_d = 350$ mm, zone plate aperture radius $R = 2.5$ mm, maximum height of the microrelief $h_{max} = 5.5$ μm, number of quantization levels $M = 256$. The values of $h_{max}$ and $M$ were chosen according to the capabilities of the technological equipment (direct laser writing system CLWS-300) used for the fabrication of the considered spectral trifocal zone plate.

As the design wavelengths, we chose the following three wavelengths approximately corresponding to the peak sensitivities of the three types of cones in the retina: $\lambda_1 = 450$ nm (blue), $\lambda_2 = 540$ nm (green), and $\lambda_3 = 580$ nm (yellow/red). The corresponding values of the refractive indices of the zone plate material are $n(\lambda_1) = 1.67$, $n(\lambda_2) = n(\lambda_3) = 1.64$ and correspond to the AZ6632 photoresist, which was used for the fabrication of the zone plate in the experiments described below. The height profile $h_{SZP}(\rho)$ of the spectral trifocal zone plate designed using Eq. (14) is shown in Fig. 1(a). It is worth mentioning that this profile contains of about 90 "irregular" height levels, which constitute a subset of the initially chosen $M = 256$ uniformly spaced levels. Let us also note that the maximum profile height of the spectral zone plate is about 5 times the height of the conventional diffractive microrelief working at a single wavelength. Typically, an increase in the $h_{max}$ value enables improving the performance of the zone plate, but makes its fabrication more complicated (mainly due to an increase in the aspect ratio).

Let the spectral multifocal lens consist of a conventional refractive lens with the focus $f_0 = 50$ mm and the designed SMZP. At the chosen values of $f_0$ and $f_d$, the following three foci are generated according to Eq. (2): $F_{-1} = 43.7$ mm, $F_0 = 50$ mm, and $F_{+1} = 58.3$ mm. Let us note that the position of the central focus can be controlled by a proper choice of the refractive lens focal length $f_0$, while the distance between the adjacent foci is defined by the $f_d$ value. Indeed, according to Eq. (2), the optical powers of the zone plate orders given by the values $P_0 = 0$, $P_{\pm 1} = \mp 1/f_d \approx \mp 2.9 \mathrm{D}$ are added to the optical power of the refractive lens $P = 1/f_0 = 20 \mathrm{D}$ thus generating 3 foci.

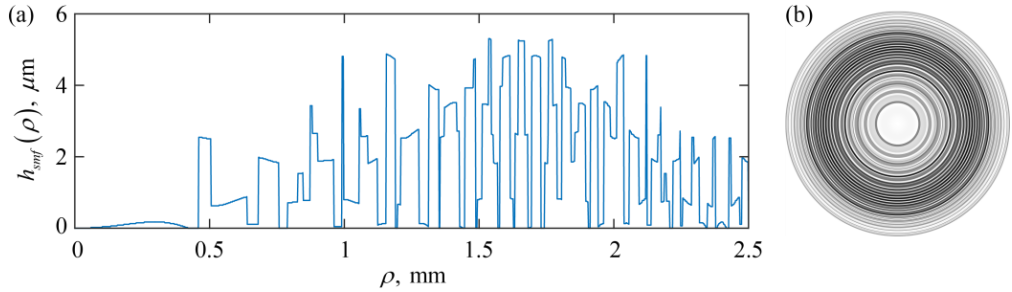

Fig. 1. Radial profile $h_{smf}(\rho)$ of the designed three-wavelength spectral trifocal zone plate (a) and the half-tone image of the zone plate microrelief used for its fabrication in resist with the laser-writing system CLWS-300 (b).

The intensity distribution generated by the spectral multifocal lens can be calculated using the Fresnel-Kirchhoff diffraction integral [1]:

$$I(r, z; \lambda) = \left| \frac{2\pi}{\lambda z} \int_0^R A(\rho) P_{smf}(\rho; \lambda) P_{lens}(\rho; \lambda) \exp\left(i \frac{\pi \rho^2}{\lambda z}\right) J_0\left(\frac{2\pi r \rho}{\lambda z}\right) r dr \right|^2, \quad (15)$$

where $z$ denotes the coordinate along the optical axis, $r$ is the radial coordinate in the plane perpendicular to the optical axis, $A(\rho)$ is the amplitude of the incident beam, $P_{smf}(\rho; \lambda) = \exp\{i(2\pi/\lambda)[n(\lambda) - 1]h_{SZP}(\rho)\}$ is the complex transmission function of the

SMZP, and $P_{lens}(\rho;\lambda) = \exp\left[-i\pi\rho^2/(\lambda f_0)\right]$ is the complex transmission function of a thin refractive lens with focal length $f_0$.

The normalized intensity distributions along the optical axis (the axial point spread function) generated by the spectral trifocal lens for the design wavelengths at $A(\rho) \equiv 1$ are shown in Fig. 2(a) and demonstrate high-quality focusing into three spots. The foci positions stay nearly the same for the design wavelengths. Let us note that the fabrication and experimental investigation of a multifocal lens containing the designed trifocal zone plate for three wavelengths is discussed below in Section 4.

According to Eqs. (12)–(14), the spectral trifocal zone plate approximates the complex transmission functions of the trifocal zone plates defined by Eqs. (1) and (5) for the given wavelengths. Let us note that the trifocal zone plate of Eqs. (1) and (5) generates uniform energy distribution between the foci at the design wavelength. Each focus can be described by the diffraction spot size $\Delta(F_m,\lambda_l) = 1.22\lambda_l F_m/R$ in the plane perpendicular to the optical axis. The values $|c_m|^2 = 0.3$, $m = 0, \pm 1$ determine the energy fractions focused in the diffraction spot of size $\Delta(F_m,\lambda_l)$ [1, 2]. It is important to note that the focal peak intensities are in inverse proportion to the squared focal lengths and the squared wavelengths, which is why the intensity values at the focal peaks are related as $I(0,F_m;\lambda_l) \sim 1/(\lambda_l F_m)^2$. The intensity distributions in Fig. 2(a) are normalized by maximum values for each of the wavelengths $\lambda_l$, $l = 1,2,3$. Thus, at the first focus (at $z = F_{-1} = 43.7$ mm) the normalized distributions reach unity and decrease at the next two foci.

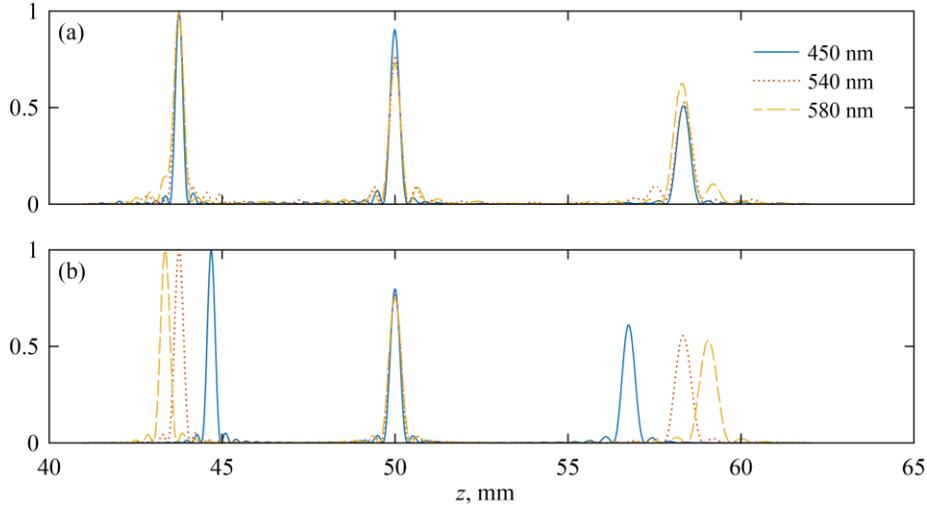

Fig. 2. The normalized intensity distributions along the optical axis generated by the spectral trifocal lens for the three design wavelengths 450 nm, 540 nm and 580 nm (a) and by the conventional trifocal lens calculated for one central wavelength 540 nm.

For comparison, Fig. 2(b) shows the normalized intensity distributions along the optical axis generated by a trifocal zone plate of Eqs. (1) and (5) calculated for a single wavelength $\lambda_2 = 540$ nm. It is evident that in this case the positions of the additional foci $F_{\pm 1}$ are not preserved for different wavelengths. For the wavelengths $\lambda_1 = 450$ nm and $\lambda_3 = 580$ nm, the distance between the foci $F_{-1}$ amounts to 1.3 mm, and the distance between the foci $F_{+1}$ is

2.3 mm. The positions of the central focus $F_0$ are identical for all the three wavelengths, since this focus is generated by the refractive lens.

## 3.2. Trifocal zone plate for the four wavelengths 450, 540, 580 and 640 nm

As a more complex example, let us consider the calculation of a trifocal zone plate with the same foci for the following four wavelengths: $\lambda_1 = 450$ nm (blue), $\lambda_2 = 540$ nm (green), and $\lambda_3 = 580$ nm (yellow/red), and $\lambda_4 = 640$ nm (yellow/red) [Fig. 3(a)]. The normalized intensity distributions along the optical axis generated by this spectral trifocal lens are shown in Fig. 3(b) and demonstrate focusing into three spots with the same positions for all the four design wavelengths. Note that as the number of the design wavelengths increases, the focal peak quality decreases, namely, the side-lobes increase and the energy distribution between the foci changes. These effects can be compensated by increasing the maximum height of the microrelief. However, as mentioned above, this will make the fabrication of such microrelief more complicated.

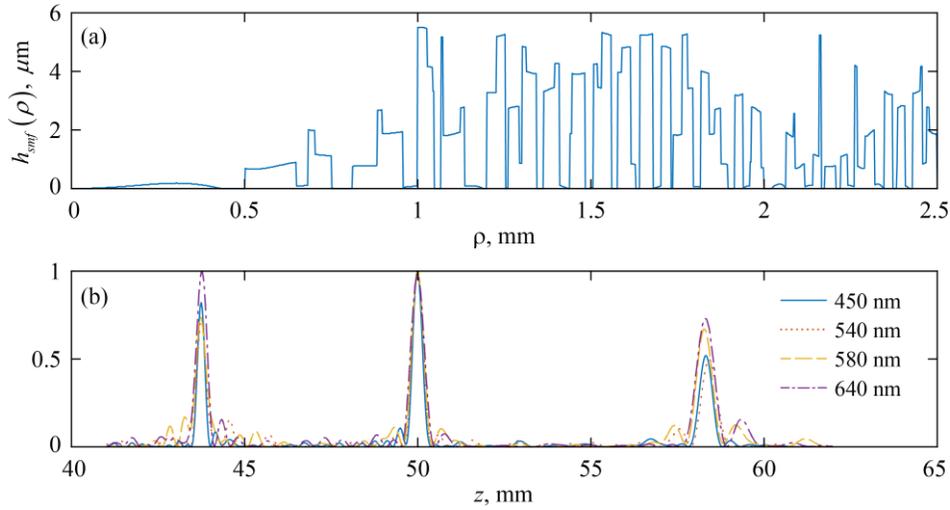

Fig. 3. (a) Radial profile $h_{smf}(\rho)$ of the four-wavelength spectral zone plate designed for the wavelengths 450 nm, 540 nm, 580 nm and 640 nm. (b) The normalized intensity distributions along the optical axis generated by the four-wavelength spectral zone plate.

## 4. Experimental results

The trifocal three-wavelength spectral zone plate described above in Subsection 3.1 [Fig. 1(a)] was fabricated by direct laser writing in photoresist using the laser writing system CLWS-300 [20]. First, a thick 6 μm layer of a positive photoresist AZ6632 was deposited on a quartz substrate. The thick layer was obtained by spin-coating two 3 μm layers, one on top of another. For the fabrication, a grayscale image corresponding to the designed zone plate was prepared [Fig. 1(b)]. The radial discretization step used in the fabrication was set to 1 μm. White areas in the image receive maximum exposure, and thus correspond to the deepest (lowest) parts of the relief. The microrelief profile of the fabricated spectral zone plate measured using a KLA-Tencor P-16+ profilometer is shown in Fig. 4(a). Figure 4(b) shows the comparison of fragments of the theoretical (blue) and measured (red) profiles. The profile of the fabricated SMZP is smoother and can be well described by the convolution of the theoretical profile with a 10 μm-wide rectangular window (green). In a future work, this model may be utilized for the compensation of the relief fabrication errors similarly to the optical proximity correction approach used in lithography. In this case, a profile has to be found, which, after the convolution with a rectangular window, approximates the theoretically calculated profile in the best way in accordance with a certain criterion.

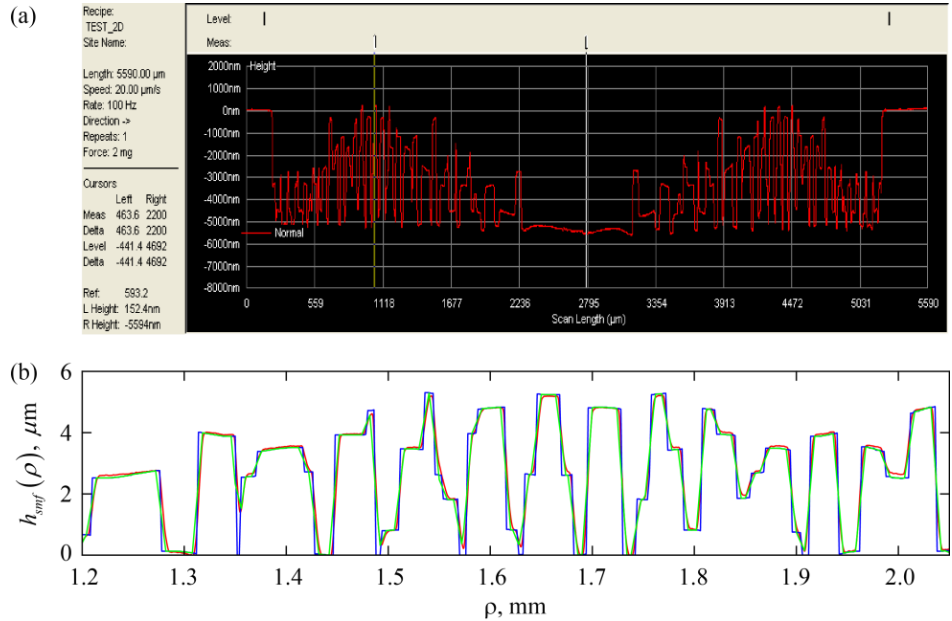

Fig. 4. (a) Measured profile of the fabricated optical element; (b) Comparison of fragments of theoretical (blue) and measured (red) profiles. The green curve corresponds to the convolution of the theoretical profile with a 10 μm-wide rectangular window.

Figure 5 shows the optical setup used for the investigation of the focusing properties of the fabricated trifocal spectral lens. As a light source 1, tunable laser Ekspla NT242 was used. The output beam was expanded up to 5 mm diameter corresponding to the SMZP diameter and collimated using two lenses 2. The obtained beam impinged on the fabricated trifocal spectral zone plate 3 placed immediately before a collecting lens 4 with focal length 50 mm. The generated intensity distributions were registered with a CCD camera 5 mounted on a motorized stage 6. The images were taken at different distances from the focusing lens in 50 μm increments. The exposure at each wavelength was adjusted so that the registered signal did not exceed the dynamic range of the used CCD sensor.

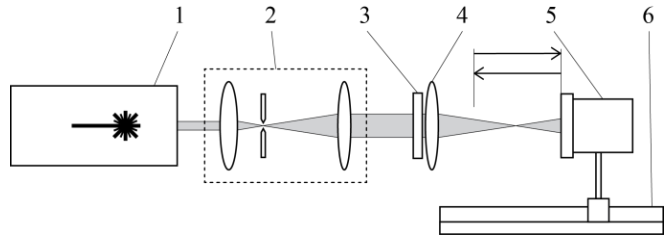

Fig. 5. Optical setup of the experiment: 1 — laser, 2 — collimator (two lenses and a 20-μm pinhole), 3 — trifocal spectral zone plate, 4 — collecting lens, 5 — CCD camera, 6 — motorized stage.

The longitudinal intensity distributions generated by the investigated lens along the optical axis for the design wavelengths are presented in Fig. 6 (blue solid lines) and demonstrate formation of three sharp foci for each of the three wavelengths. Similarly to Fig. 2(a), the intensity distributions in Fig. 6 are normalized by maximum values for each of the wavelengths $\lambda_l$, $l = 1, 2, 3$. Note that the lens 4 and the collimating lenses 2 possess chromatic aberrations affecting the position of the central focus. Thus, to compare the positions of the additional foci with respect to the central focus, the longitudinal intensity distributions in Fig. 6 were shifted so that the central focus in all the three cases corresponded to $z = 50$ mm. Taking into account this correction of the central focus positions, the positions of the additional focal peaks are almost the same for the design wavelengths and are in good agreement with the peaks of the theoretical distributions (shown in Fig. 6 with red dashed lines). At the same time, there are some discrepancies between the calculated and registered intensity distributions, namely, the registered peaks are broader and have different relative intensities than in the theoretically calculated distributions. The observed discrepancies result from the spherical aberrations of the refractive spherical lenses used in the experimental setup [9], as well as from the defects in the profile of the fabricated spectral multifocal zone plate.

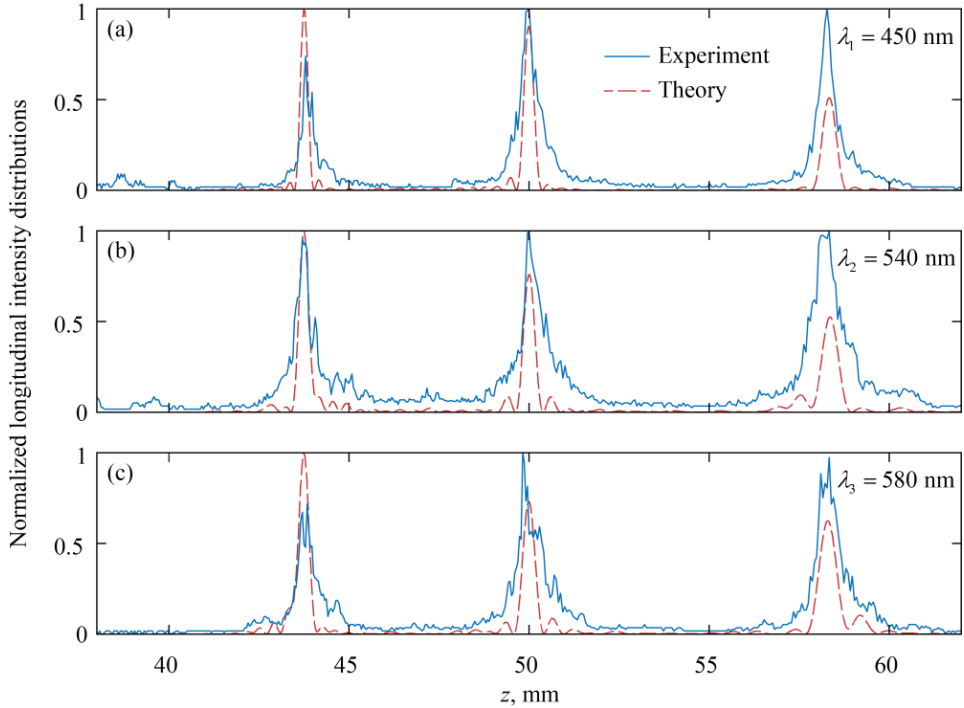

Fig. 6. Normalized registered longitudinal intensity distributions generated by the fabricated lens (blue solid lines) for the three design wavelengths 450 nm (a), 540nm (b), and 580 nm (c), and the normalized theoretical distributions calculated for the spectral trifocal lens of Fig. 1(a) (red dashed lines).

The transverse intensity distributions measured at the three focal planes (at the three local maxima of the longitudinal intensity distributions) at the design wavelengths $\lambda_1 = 450$ nm, $\lambda_2 = 540$ nm, and $\lambda_3 = 580$ nm are shown in Figs. 7(a), 7(b) and 7(c), respectively. Figure 7 confirms the formation of three sharp foci for each of the design wavelengths. The intensity distributions in Fig. 7 were normalized in the same way as the longitudinal distributions shown in Fig. 6. Red circles in Fig. 7 depict the theoretical diffraction spot sizes with the diameter calculated according to the expression $\Delta(F_m, \lambda_l) = 1.22 \lambda_l F_m / R$, where $l = 1, 2, 3$, $m = -1, 0, 1$. For illustrative purposes, these circles are shown not in the zero-intensity plane, but in the plane corresponding to the normalized intensity value of 0.6. The sizes of the measured peaks are in good agreement with the theoretical values. Again, a certain increase in the size of the measured peaks can be attributed to the spherical aberrations of the used refractive spherical lenses and to the fabrication defects.

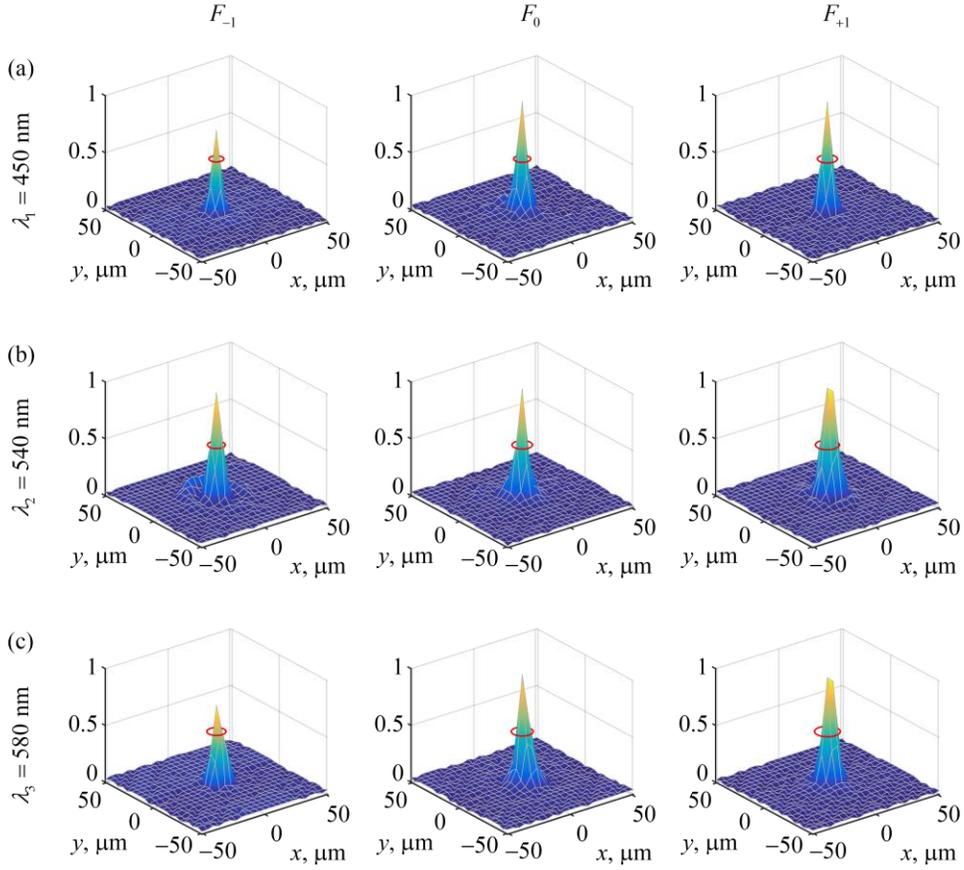

Fig. 7. Normalized registered transverse distributions generated by the fabricated lens at the three design wavelengths 450 nm (a), 540 nm (b), and 580 nm (c) at the three foci. Red circles depict the theoretical diffraction spot sizes.

## 5. Conclusion

A method for designing diffractive spectral multifocal lenses generating a set of fixed-position foci at several different wavelengths has been proposed. The presented method is based on minimizing the difference between the complex amplitudes of the beams generated by the microrelief of the spectral lens at the design wavelengths $\lambda_l$, $l = 1,...,L$, and the complex transmission functions of "monochromatic" multifocal diffractive lenses calculated for these wavelengths.

As an example, a trifocal zone plate with 5 mm diameter and optical powers $P_{-1} = 3\text{D}$, $P_0 = 0$, and $P_{+1} = -3\text{D}$ operating at three wavelengths $\lambda_1 = 450$ nm, $\lambda_2 = 540$ nm, and $\lambda_3 = 580$ nm has been designed and fabricated using direct laser writing in photoresist. The results of proof-of-concept experimental investigations of the trifocal spectral lens consisting of a conventional refractive lens and the fabricated spectral zone plate have confirmed the formation of three fixed-position foci for the design wavelengths. The obtained results may find applications in the design and fabrication of novel multifocal intraocular lenses with reduced chromatic effects.

**Funding**

Ministry of Education and Science of Russian Federation (theoretical part of the work); Russian Foundation for Basic Research (RFBR) (16-29-11683) (experimental part of the work including the microrelief fabrication).